# On the validity of using idealised sample geometries for interpreting mechanical tests of very soft tissues

Sajjad Arzemanzadeh, Karol Miller, Adam Wittek

Intelligent Systems for Medicine Laboratory, The University of Western Australia, Perth, Western Australia, Australia

## Abstract

Mechanical characterisation of soft tissues often relies on inverse analysis of experimental data in which constitutive models are calibrated to match experimental force–displacement curves, yet the vast majority of such studies use idealised (nominal) sample geometries even though experimental samples unavoidably deviate from these nominal shapes because of imperfections in excision and mounting. The influence of these geometric simplifications on the material parameters determined through inverse analysis remains poorly quantified. We investigate the appropriateness of using idealised sample geometries in mechanical characterisation of brain tissue. Magnetic resonance imaging (MRI) was used to reconstruct the exact (real) geometry of each nominally cuboidal tissue sample. We determined a stress parameter (the shear modulus) by modelling, using the finite element method, tensile, compressive, and shear tests of brain tissue samples with both the MRI-based (real) and idealised cuboidal geometries, enabling a controlled comparison of geometry. Idealised geometries consistently yielded a lower stress parameter. The discrepancy in shear modulus between the real and idealised geometries varied across loading modes, averaging approximately 10% in shear and 48% under axial loading, predominantly arising from the compressive response. These discrepancies can be attributed to the inability of idealised-geometry models to accurately represent contact interactions and predict strain distributions, particularly under compressive loading. Idealisation of sample geometry may introduce systematic bias in the mechanical characterisation of very soft tissues; therefore, the actual measured sample geometry should be used in inverse analysis to identify constitutive models and their parameters.

**Keywords:** Very soft tissue, Brain tissue, Mechanical properties, Sample geometry, Finite element method

## 1. Introduction

A Google Scholar search for *"soft tissue biomechanics experiment"* returns approximately 179,000 documents. Despite this extensive body of research, recent literature reviews report inconsistencies and even contradictions in experimentally derived mechanical properties of very soft tissues, including the brain [1-3] and liver [4, 5]. Mechanically characterising very soft tissues (and other very soft materials) is inherently difficult: their extreme compliance challenges conventional test configurations designed for stiffer materials and has prompted a range of specialised experimental techniques for measuring their properties [6-9]. Therefore, Experimental procedures for characterising very soft tissues typically follow those used for engineering materials: material samples are subjected to imposed displacements and resulting forces are measured. These measurements, combined with sample geometry, are used to infer a constitutive model and its parameters using inverse analysis. However, experimental investigation of very soft tissues typically involves large deformations, often exceeding 30% strain, such that changes in sample geometry during testing must be accounted for. This necessitates the use of geometrically non-linear formulations based on finite deformation theory. When combined with the intrinsically non-linear constitutive behaviour of very soft tissues, these factors render closed-form analytical solutions challenging and not always practical. Consequently, methods of computational mechanics, most commonly the finite element (FE) method, are employed to model very soft tissue mechanical tests and conduct inverse identification of tissue material properties [10-16].

Unlike analytical methods, which require simple geometries such as cylinders or cuboids, FE discretisation allows accurate representation of arbitrary sample geometries. However, only a small number of studies have attempted to capture and use the actual three-dimensional geometry of very soft tissue samples using camera-based methods [17], laser scanning [12], and radiographic imaging [18, 19].

Vast majority of studies assume nominal, idealised cuboidal [20, 21] or cylindrical [11, 22-25] geometries. This is despite the well-known difficulties associated with excising brain tissue specimens that would conform to such ideal shapes. Regardless of the excision/"cutting" technique employed, whether specialised microtome blades [24, 25], scalpels [21, 26], or punches [22, 23, 27-29], the resulting specimen surfaces typically lack flatness, and the actual geometries inevitably deviate from their nominal shapes. These deviations are particularly pronounced for very soft tissues such as the brain. For instance, Budday et al. reported edge lengths ranging from 3 to 7 mm and a height ranging from 2 to 5 mm for cuboidal samples of the brain tissue with nominal dimensions of 5 mm × 5 mm × 5 mm [20].

Despite this, the influence of discrepancies between actual and idealised sample geometry on the modelling of mechanical tests and on the material properties derived from such tests has not yet been quantified. The present study addresses this gap by investigating the appropriateness of using idealised sample geometries in modelling very soft tissue mechanical testing. We quantitatively compare ovine brain tissue stress parameter (shear modulus) in tension, compression, and shear obtained using idealised geometries with those derived from sample-specific three-dimensional geometries reconstructed from high-resolution magnetic resonance imaging (MRI).

## 2. Materials and Methods

### 2.1. Sample Preparation

All experiments were conducted on ovine brain tissue samples extracted from fresh skinned sheep cadaver heads. Sample preparation, MRI, and biomechanical testing were completed within 24 hours post-mortem to minimise the risk of alteration of mechanical properties. This timeframe is shorter than post-mortem delays known to significantly alter the mechanical characteristics of the brain tissue [30]. The skull bone was carefully removed, and one or two brain tissue samples were excised from each head from the same region. We used a microtome blade for cutting the brain tissue to obtain nominally cuboidal samples, approximately 23 mm × 20 mm × 13 mm. After excision, the bottom surface of the samples was immediately attached to custom 3D-printed platens using a fast-curing cyanoacrylate glue to facilitate handling without direct contact with the tissue and minimise the risk of unintentional sample damage. To maintain hydration during handling, the samples were placed in a sealed container containing approximately 10 mL of 5% physiological saline solution, without direct contact with the liquid, and were kept moist by exposure to saline vapour. In total, we prepared and tested 15 samples: five under shear loading, five under tension, and five under compression.

### 2.2. Acquisition of Sample Images

MRI scans of all samples were acquired before mechanical testing. For tension and compression samples, a Bruker Biospec 9.4T scanner with a 3D T2-weighted TurboRARE sequence was used, resulting in a resolution of 0.25 mm × 0.25 mm × 0.5 mm. For shear samples, a Siemens Healthineers Vida 3T scanner with a 2D T2-weighted turbo spin echo sequence was used, resulting in a resolution of 0.6 mm × 0.6 mm × 0.6 mm. A different MRI scanner was used for shear tests due to an unexpected failure of the 9.4T scanner; however, the T2-weighted sequence was carefully matched to that used for tension and compression samples.

### 2.3. Mechanical Testing

**Fig. 1** shows the experimental setups. The UniVert (CellScale) specialised biomaterial testing system equipped with a 2.5 N load cell (accuracy of 0.1% of the maximum load) was used to conduct mechanical tests. The force and loading platen displacement were recorded with a sampling rate of 100 Hz. The experimental protocol follows the general guidelines of Miller [6, 7] and Miller and Chinzei [22, 23]. The 3D-printed plates used to handle the samples were rigidly attached using adhesive to the base platen of the testing machine.

No-slip boundary conditions between the sample top surface and UniVert testing system loading platen were created. In compressive tests, it was achieved by attaching an 80-grit sandpaper to the loading platen. In tensile and shear tests, the sample top surface was glued to the loading platen using fast-curing cyanoacrylate glue. To ensure adhesion, after the initial contact of the sample top surface with the loading platen, the sample was compressed by 1 mm for 30 s. It was then returned to the original unloaded configuration and held there for an additional 180 s to provide stress relaxation before starting the tests. The time of 180 s was selected as it has been reported in the experimental literature as sufficient for brain tissue relaxation [31]. The loading speed was set to 0.3 mm/s, which implies a nominal strain rate of approximately 0.02 $s^{-1}$. All tests were conducted at room temperature. The experiments were recorded using three digital cameras, which provided the frontal and oblique views of the sample (**Fig. 1**).

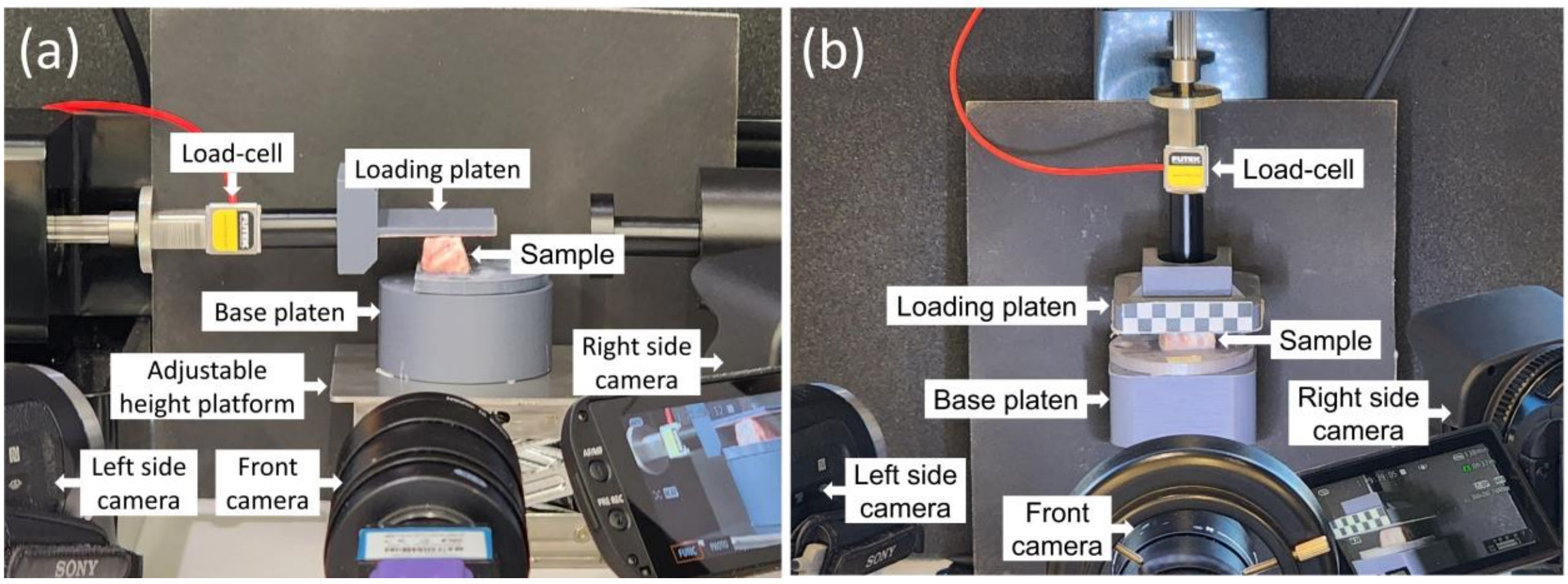

**Fig. 1** Experimental setup for (a) shear test and (b) tensile and compressive tests on brain tissue samples

### 2.4. Finite element models for experiment simulation

#### 2.4.1. Geometry Reconstruction

The acquired MRIs were imported into 3D Slicer, free, open-source software for analysis of medical, biomedical, and other 3D images [32], to extract accurate three-dimensional sample geometries. This was achieved through semi-automatic segmentation using the thresholding and Gaussian smoothing tools within the 3D Slicer Segment Editor module [33]. These segmentations served as the reference "real geometry" to construct computational grids (FE meshes) for subsequent FE analyses. Idealised cuboidal geometry for each sample was obtained by averaging the dimensions measured using the 3D Slicer Markups module, at the top, middle, and bottom cross-sections of the segmented sample MRIs. Unlike measurements obtained using contact devices such as callipers and micrometres, which are prone to inter- and intra-operator variability, this approach enables consistent comparison with the real segmentation-based geometries (see **Fig. 2**).

#### 2.4.2. Finite Element Mesh Generation and Element Formulation

All finite element meshes were generated using Coreform Cubit meshing software [34]. We used 8-noded under-integrated hexahedral elements (C3D8R element type in Abaqus finite element code) with stiffness-based hourglass control [35]. This element does not exhibit volumetric locking which makes it appropriate for modelling of the brain tissue and other nearly incompressible continua. An element size of 0.5 mm was used for all brain tissue samples as it is consistent with the resolution of acquired MRIs and ensures mesh-independent converged solution. All simulations were conducted using the established Abaqus explicit dynamics non-linear FE solver with automated time stepping [35]. Time increments of approximately $2.5 \times 10^{-5}$ s ensured stability and accuracy of this explicit scheme [35].

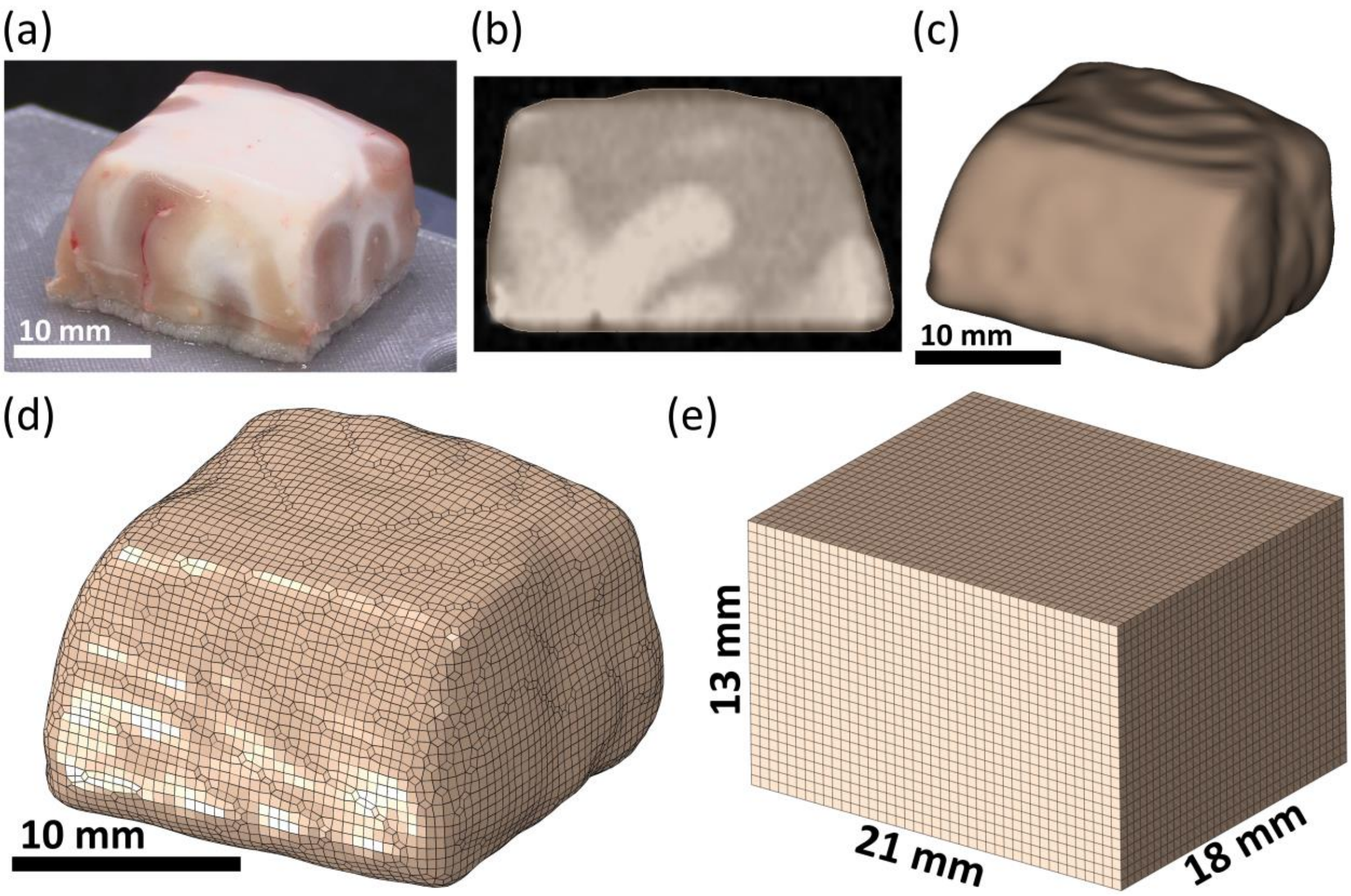

**Fig. 2** (a) A typical brain tissue sample used in this study, (b) the MRI of the sample with highlighted material segmentation, (c) 3D representation of the segmentation, (d) meshed geometry imported into Abaqus, and (e) idealised geometry counterpart with sample-specific dimensions. In this example, there are 40633 elements in the real geometry and 36244 elements in the idealised geometry

#### 2.4.3. Boundary Conditions and Loading

Boundary conditions were defined to replicate the experimental setup (**Fig. 3**). For all loading modes, nodes of the bottom surface of the samples were fully constrained. For tensile and shear tests, the loading was applied to the top surface nodes via a prescribed displacement. Nodes for prescribing the displacements were identified from the test video recordings. For compressive tests, the loading and base platen were modelled by rigid plates, and no-slip contact algorithm from the Abaqus finite element code was used to represent the interactions between the brain tissue sample top surface and moving loading platen. The time histories of nodal displacements and loading platen movement were defined using the smooth step procedure from Abaqus FE code [35] which utilises a fifth-order polynomial.

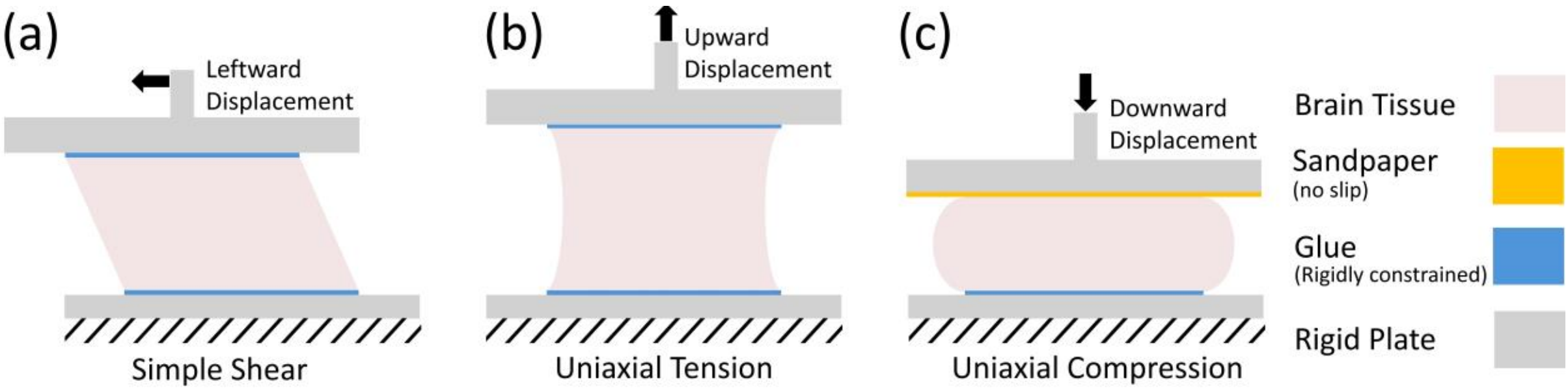

**Fig. 3** Schematic of the boundary and loading conditions for (a) simple shear, (b) uniaxial tensile, and (c) uniaxial compressive tests conducted in this study

### 2.4.4. Material Model

In recent years, neural networks have been increasingly applied to soft-tissue mechanical characterisation for identifying the form of the strain energy function or constitutive model that best represents tissue mechanical responses [36-39]. However, we did not adopt this powerful approach, as the aim of the present study is considerably narrower than full constitutive model discovery. As over the past thirty years, the Ogden constitutive model has been extensively used in brain biomechanics [22, 39, 40], we modelled the mechanical behaviour of the brain tissue using a second-order Ogden [41] hyperelastic constitutive law:

$$W(\overline{\lambda_1}, \overline{\lambda_2}, \overline{\lambda_3}) = \sum_{i=1}^{N} \frac{2\mu_i}{\alpha_i^2} \left( \overline{\lambda_1}^{\alpha_i} + \overline{\lambda_2}^{\alpha_i} + \overline{\lambda_3}^{\alpha_i} - 3 \right) + \sum_{i=1}^{N} \frac{1}{D_i} (J_{el} - 1)^{2i}, \quad (1)$$

where $\overline{\lambda_\iota}$ are deviatoric principal stretches, $J_{el}$ is the elastic volume strain, $\mu_i$ are the shear parameters, $\alpha_i$ are the dimensionless constants, $D_i$ are material constants determining compressibility, and $N$=2 for second-order Ogden model. The shear modulus is given by

$$\mu = \sum_{i=1}^{N} \mu_i, \quad (2)$$

and the initial bulk modulus is given by

$$K_0 = \frac{2}{D_1}. \quad (3)$$

As the brain is nearly incompressible, we used a Poisson's ratio of 0.49 [42]. A unique set of material constants ($\mu_i$ and $\alpha_i$) was determined for a strain of up to 0.25 for tensile tests, and 0.3 for compressive and shear tests. These strain limits are consistent with literature indicating that deformations exceeding these limits can induce irreversible mechanical damage in brain tissue [6, 7].

### 2.5 Determining Material Constants

For each sample, we determined the brain tissue shear parameters $\mu_i$ and the exponents $\alpha_i$ (Equation 1) using an inverse FE approach formulated as an optimisation problem solved using the Sequential Least Squares Programming (SLSQP) algorithm [43] implemented in SciPy library in Python. We defined the objective function as the absolute area between the experimental and FE-predicted force-displacement curves, necessitating a FE simulation for each evaluation of the objective function. Optimisation space was bounded by the biomechanically plausible values of shear parameters $\mu_i$ and $\alpha_i$. The optimisation was carried out separately for the models built using tissue sample geometries obtained directly from the MRIs and idealised cuboidal geometries, enabling quantitative assessment of the influence of geometry simplification on the identified material constants. Because soft tissues exhibit different stiffness under tension and compression, a single axial loading mode (either tension or compression) is insufficient to determine their material properties [22]. Therefore, the calibration was done simultaneously for compressive and tensile tests on samples extracted from the same sheep brain.

## 3. Results

### 3.1. Geometry Reconstruction

Cross-section areas of the sample bottom surfaces glued to the baseplate showed small differences between real and idealised geometries, with the average difference of 6.5% (SD of 4.7%) across all samples. However, for the area of the sample top surface, where the load was applied, greater differences and variation were observed. On average, this area was smaller for real geometries than for idealised cuboidal geometries by 15.6% (SD of 2.5%) in shear tests, 19.8% (SD of 7.5%) in tensile tests, and 30.2% (SD of 4.1%) in compressive tests.

### 3.2. Determining Material Constants

The stress–strain curves obtained from the experiments conducted in this study are consistent with the literature on the brain tissue mechanical behaviour, showing higher nominal stress under compression compared to tensile and shear (**Fig. 4**).

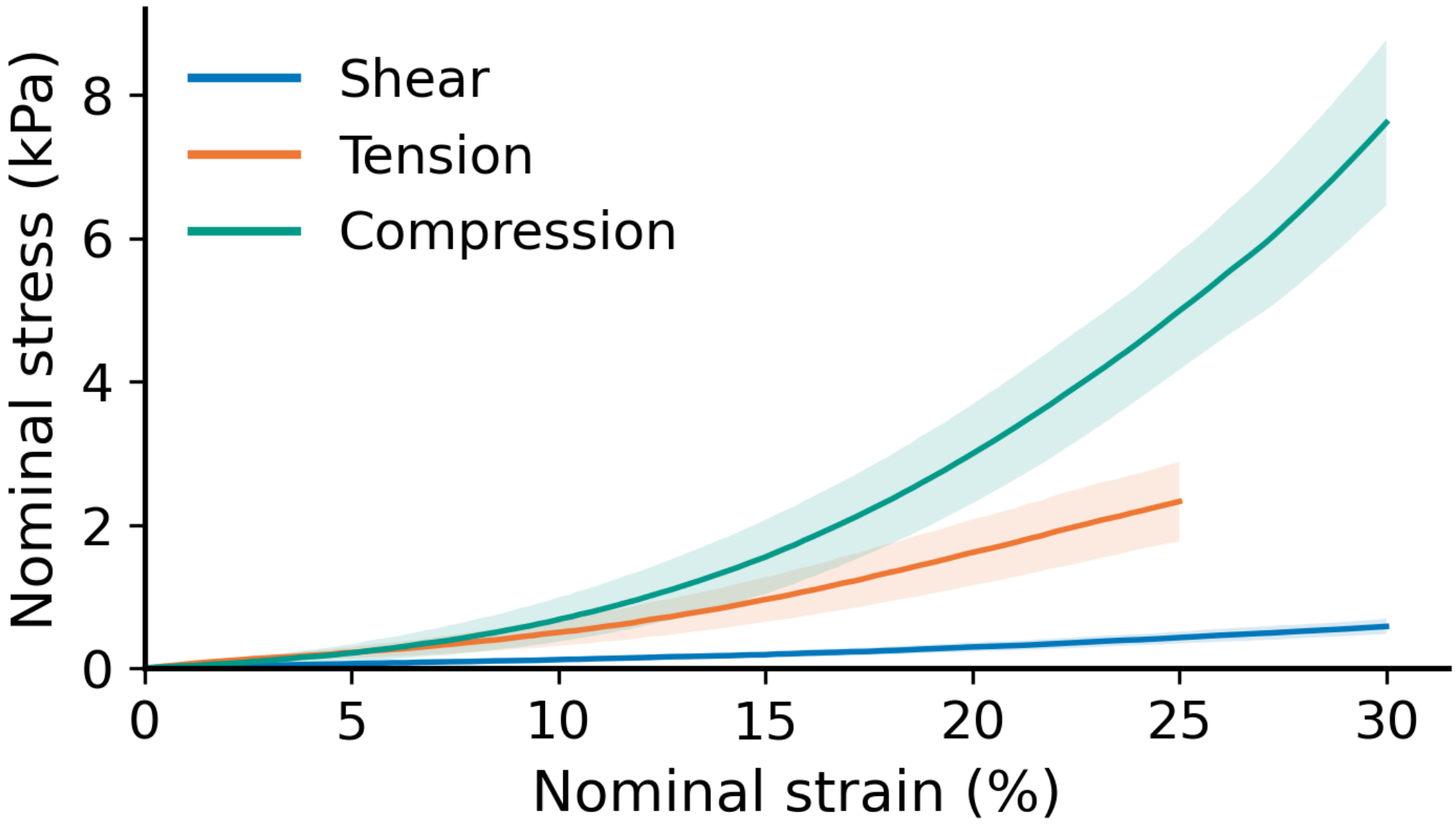

**Fig. 4** Nominal stress-strain curve of the brain tissue under shear, tension and compression experimentally obtained in this study. To enable comparison between different loading modes, absolute stress and strain values are used (i.e. compressive stress and strain are reported with positive values). Solid lines represent the nominal stress averaged over five samples under each loading mode (shear, tension, and compression – see **Fig. 5**). The shaded bands around the solid curves indicate +/- standard deviation SD bounds

Initial four-parameter optimisation indicated that the exponent factors remained effectively constant across different samples and loading conditions, with $\alpha_1$ consistently close to −8 and $\alpha_2$ close to 16. Consequently, we performed a two-parameter optimisation, calibrating only the shear parameters $\mu_1$ and $\mu_2$, while holding $\alpha_1 = -8$ and $\alpha_2 = 16$ constant.

FE models using both real and idealised geometries accurately represented experimental force–displacement characteristics measured in shear, tensile and compressive tests (**Fig. 5**), indicating the appropriateness of the selected material model, our framework for determining material constants, and definition of boundary conditions in the models. However, the shear modulus determined from the models using idealised cuboidal geometries was consistently lower than those from the models using the real sample geometries. The magnitude of this trend varied depending on the loading type.

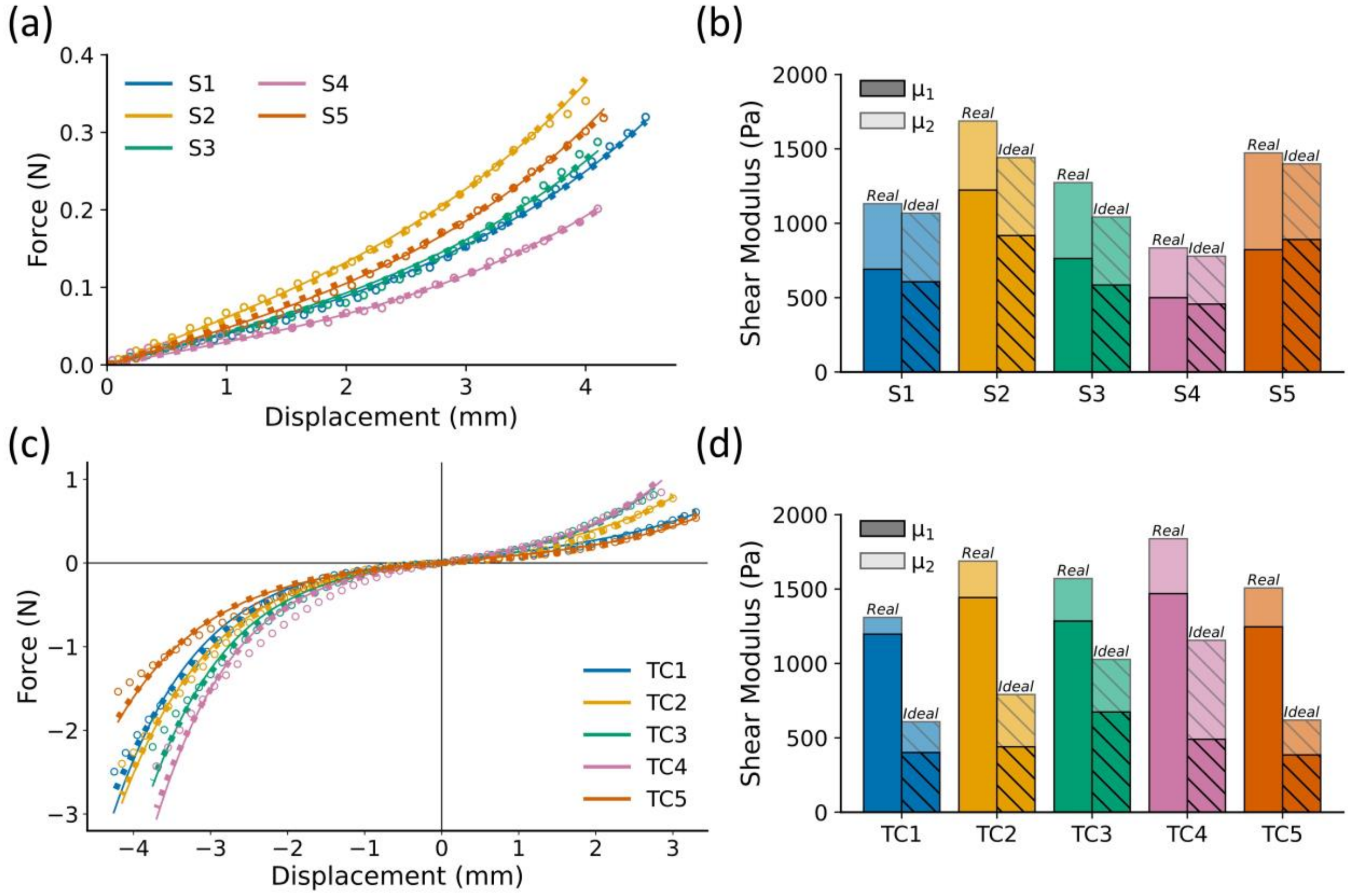

**Fig. 5** Results obtained using second-order Ogden constitutive model, with $\alpha_1 = -8$ and $\alpha_2 = 16$, to describe the mechanical behaviour of the brain tissue under shear, and tension and compression. Force-displacement curves obtained from the experiments and FE models of (a) shear and (c) tensile and compressive tests on brain tissue samples. Experimental data are plotted via hollow circles (○) showing only every fiftieth data point for clarity. Results from the FE models using real sample geometry are plotted via solid lines (–). Results from the FE models using idealised sample geometry are plotted via dotted lines (---). Calibrated $\mu_1$ and $\mu_2$ using real and idealised geometries for (b) shear and (d) tensile and compressive tests on brain tissue samples. $\mu_1$ and $\mu_2$ are coloured in dark and light contrasts respectively. Colour codes are consistent between the right and left columns for each row. Shear tests, each conducted using a tissue sample extracted from different sheep cadaver brain, are labelled as S1 through S5. Tensile and compressive tests, conducted using two tissue samples extracted from the same sheep cadaver brain, are labelled together as TC1 through TC5

**Table 1** Calibrated shear modulus of brain tissue using a second-order Ogden constitutive model, with $\alpha_1 = -8$ and $\alpha_2 = 16$, obtained from the FE models using real (determined from the MRIs) and idealised (cuboidal) tissue sample geometries. For tension and compression, the calibration was done simultaneously using the experimental results obtained for the samples extracted from the same sheep cadaver brain. $\mu_1$ and $\mu_2$ are the shear parameters and $\mu = \mu_1 + \mu_2$ is the shear modulus. Difference is defined as $\frac{\mu_{\text{Ideal}} - \mu_{\text{Real}}}{\mu_{\text{Real}}} \times 100$

| Loading | Sample(s) | Second-order Ogden: $\alpha_1 = -8$ and $\alpha_2 = 16$ | | | | | | |
|---|---|---|---|---|---|---|---|---|
| | | Real | | | Idealised | | | Difference (%) |
| | | $\mu_1$ | $\mu_2$ | $\mu$ | $\mu_1$ | $\mu_2$ | $\mu$ | |
| Shear | S1 | 691.0 | 440.4 | 1131.4 | 606.6 | 461.6 | 1068.2 | -5.6 |
| | S2 | 1222.6 | 464.3 | 1686.9 | 917.3 | 523.7 | 1441.0 | -14.6 |
| | S3 | 763.0 | 510.0 | 1273.0 | 584.9 | 456.8 | 1041.7 | -18.2 |
| | S4 | 499.8 | 334.4 | 834.2 | 456.9 | 322.0 | 778.9 | -6.6 |
| | S5 | 822.0 | 651.0 | 1473.0 | 890.9 | 508.2 | 1399.1 | -5.0 |
| Tensile and Compressive | TC1 | 1196.8 | 113.1 | 1309.9 | 401.2 | 207.3 | 608.5 | -53.5 |
| | TC2 | 1443.0 | 245.3 | 1688.3 | 439.7 | 351.5 | 791.2 | -53.1 |
| | TC3 | 1284.6 | 286.1 | 1570.7 | 673.7 | 353.8 | 1027.5 | -34.6 |
| | TC4 | 1469.4 | 368.8 | 1838.2 | 490.5 | 665.4 | 1155.9 | -37.1 |
| | TC5 | 1246.5 | 261.2 | 1507.7 | 385.4 | 234.0 | 619.4 | -58.9 |

For shear loading, the discrepancy between the shear modulus determined using the real and cuboidal geometries was the lowest, with the idealised geometries yielding shear modulus on average 10% (SD of 6%) lower than those determined using the real geometries.

For tensile and compressive loading, the effects of tissue sample geometry simplifications were more pronounced than for shear, with the shear modulus from the models using the idealised cuboidal geometry on average 47.5% (SD of 10.9%) lower than that from the models using the real geometry. In our second-order Ogden model, $\mu_1$ associated with $\alpha_1 = -8$ mainly governs the compressive and very low strain tensile behaviour, while $\mu_2$ associated with $\alpha_2 = 16$ is more dominant beyond low strain tensile region. From **Table 1**, the difference between $\mu_1$ obtained from the real and idealised geometry is substantially larger than the difference for $\mu_2$, which suggests that the results for compressive loading are the main source of the difference in calibrated $\mu$ between the real and idealised geometry. To evaluate this hypothesis, we applied a first-order Ogden model to compressive tests, with the shear modulus calibration using the real and idealised geometry. As shown in **Fig. 6** and reported in **Table 2**, simplifying the sample geometry as a cuboid resulted in shear modulus that was on average 48.5% (SD of 11%) lower than that obtained using the real sample geometry (**Fig. 6** and **Table 2**). This difference is very close to that of 47.5% observed when the calibration was done simultaneously for tension and compression using a second-order Ogden model, which confirms that the difference between shear modulus determined using the real and idealised geometry is determined by compression rather than tension. This effect appears to be because unlike idealised models, the actual brain tissue samples did not maintain full contact with the loading platen for a significant portion of each test. This incomplete contact was due to surface imperfections, including non-planarity, misalignment (i.e. parts of the top sample surface were not parallel to the loading platen), and the presence of localised depressions and peaks (see **Fig. 2** and **Fig. 7**). Although these imperfections are clearly visible in the sample MRIs (**Fig. 2b, c**), they may not be readily detected through visual inspection or photography that create the impression of smooth surfaces (**Fig. 2a**). In contrast, models employing idealised cuboidal geometries imply full and uniform contact throughout loading.

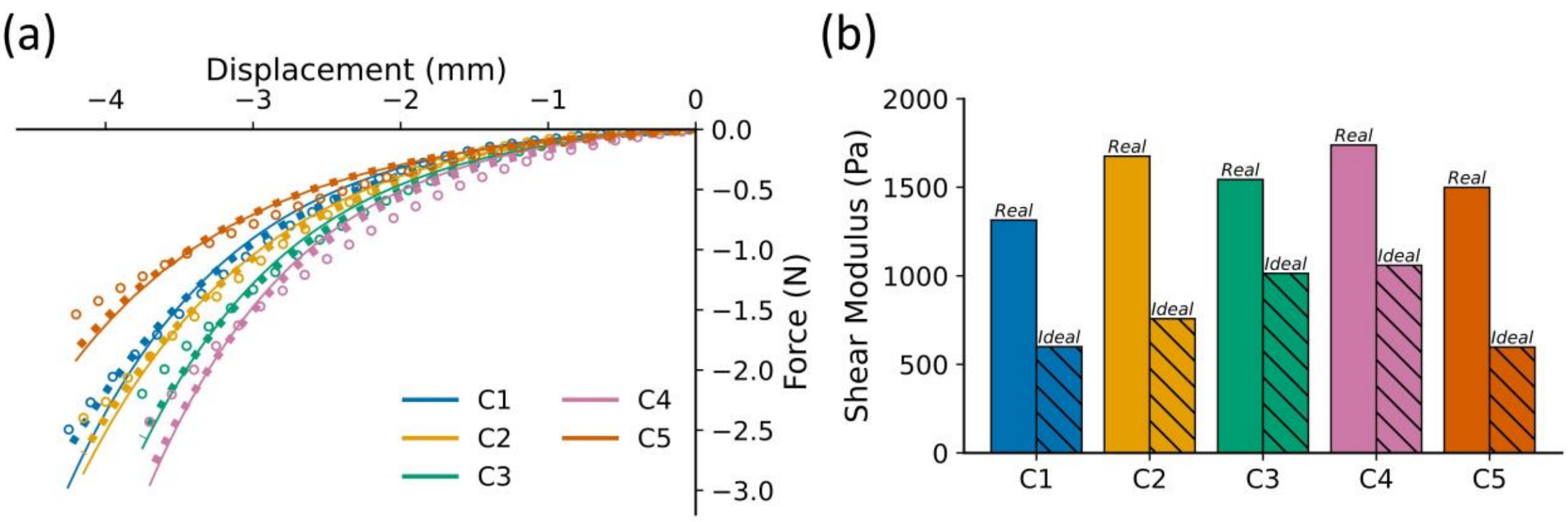

**Fig. 6** Results obtained using first-order Ogden constitutive model with $\alpha = -8$ to describe mechanical behaviour of the brain tissue under compression. (a) Force-displacement curves obtained from the experiments and FE models of the experiments. Experimental data are plotted via hollow circles (○) showing only every fiftieth data point for clarity. Results from the FE models using real sample geometry are plotted via solid lines (–). Results from the FE models using idealised sample geometry are plotted via dotted lines (---). (b) Calibrated shear modulus $\mu$ obtained using real and idealised geometries for compressive tests. Colour codes are consistent between the right and left columns for each row

**Table 2** Calibrated shear modulus of brain tissue using a first-order Ogden constitutive model, with $\alpha_1 = -8$, obtained from the FE models using real (determined from the MRIs) and idealised (cuboidal) tissue sample geometries. Difference is defined as $\frac{\mu_{Ideal} - \mu_{Real}}{\mu_{Real}} \times 100$

| Loading | Sample | First-order Ogden: $\alpha = -8$ | | |
|---|---|---|---|---|
| | | Real | Idealised | Difference (%) |
| | | $\mu$ | $\mu$ | |
| Compressive | C1 | 1314.5 | 599.4 | -54.4 |
| | C2 | 1675.4 | 757.9 | -54.8 |
| | C3 | 1543.9 | 1013.5 | -34.4 |
| | C4 | 1738.5 | 1059.2 | -39.1 |
| | C5 | 1499.1 | 597.5 | -60.1 |

Here, we used rigid analytical plates [44] to model the loading platens in compressive tests, whereas some studies apply prescribed displacements directly to their FE model nodes. While this approach may be acceptable for idealised geometries at compressive strains up to approximately 15%, it becomes inappropriate at higher strains, as contact between the loading platen and the sample side walls is neglected [15]. This limitation is more pronounced for real geometries, where accurate modelling of the progressive development of contact is essential and cannot be captured by direct nodal displacement.

The geometric imperfections and partial contact with the loading platen led to non-uniform and non-symmetric deformation and strain distribution within the brain tissue samples during compression (**Fig. 7c**). Consequently, the deformation patterns observed experimentally could not be accurately captured by models using cuboidal geometries, in contrast to those using real sample geometries reconstructed from the MRIs. Localised strain variations predicted by the models using real geometries are absent in the cuboidal models (**Fig. 7**).

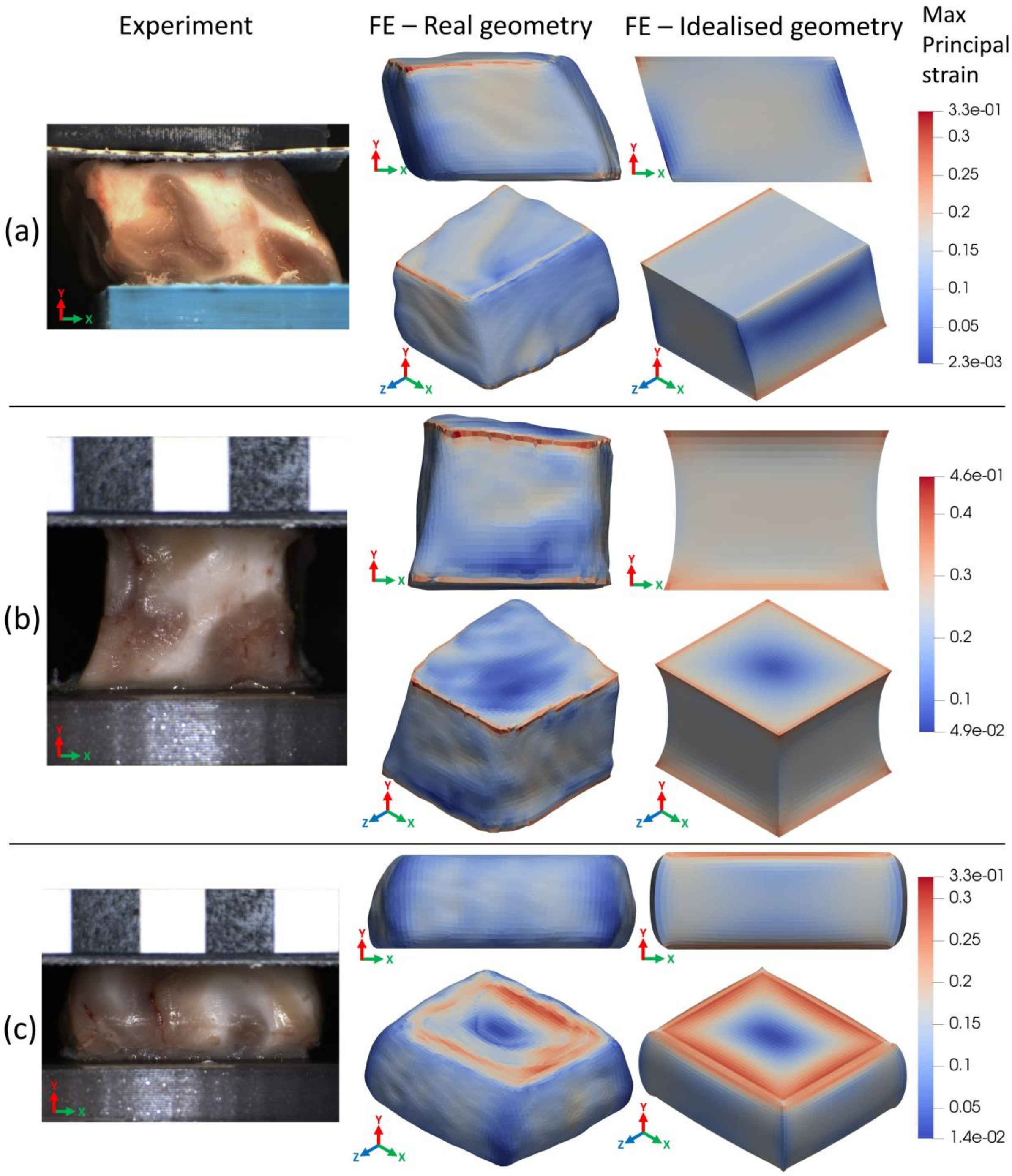

**Fig. 7** Brain tissue samples under (a) shear loading at 0.3 nominal strain, (b) tensile loading at 0.25 nominal strain, and (c) compressive loading at 0.3 nominal strain. The first column is the front view of the tissue sample, the second and third columns are maximum principal Hencky strain maps of FE models using the real geometry and idealised geometry, respectively. Faces of the tissue samples are not flat (exhibit imperfections) which is reflected in the non-uniform strain distribution obtained from the models using real geometry

## 4. Discussion

We examined how simplifying sample geometry affects determination of brain tissue stress parameter (shear modulus) using mechanical testing combined with inverse finite element (FE) analysis. Real (or true) sample geometries for FE models were reconstructed from segmented magnetic resonance images (MRIs), capturing surface imperfections and deviations from the nominal cuboidal shape. For comparison, idealised cuboidal geometries were created by averaging edge lengths measured from the MRI segmentations, producing cuboids closely matching those of the real geometries and enabling controlled evaluation of geometric simplification effects.

Our results indicate that incorporating realistic sample geometries leads to consistently higher shear modulus compared with that obtained from idealised cuboidal geometries — approximately 10% higher under shear loading, and about 48% higher under axial loading, with the discrepancy predominantly arising from the compressive response. One important reason for these differences is that idealised models assume fully developed contact across the entire top surface of the sample and the loading platen from the very start of the experiment. This assumption substantially overestimates the true contact area. Because a larger contact area generates a higher reaction force for a given material stiffness, the optimisation algorithm compensates by reducing the shear modulus to match the experimentally measured forces. This effect is particularly significant in compression, where the difference in the shear modulus obtained from the models using real and cuboidal geometries approaches 50%. In both the experiments and the finite element models using realistic sample geometries, contact between the loading platen and the sample top surface developed locally at first rather than uniformly over the entire surface. The contact area increased progressively with compressive deformation. This behaviour gives rise to a more complex and non-symmetric strain distribution compared with the idealised cuboidal models, which assume a perfectly flat surface with full contact from the loading onset. These differences in predicted strain fields are a major factor underlying the large discrepancies in stress parameters determined under compressive loading when comparing real and idealised geometries.

It should be noted that, although this study is one of the few investigations that confirm and emphasise the importance of accurately representing sample geometry when determining the mechanical properties of very soft tissues through combined experiments and inverse finite element analysis, the challenge of accounting for geometric imperfections has long been recognised in the engineering literature, particularly in relation to predicting the behaviour of structures under compression [45-49].

We used nominally cuboidal brain tissue samples. Nominally cylindrical samples are also commonly employed. However, geometric imperfections similar to those observed here have also been reported for cylindrical very soft tissue samples [23]. Similar effects are therefore likely to occur also for nominally cylindrical samples.

## 5. Conclusions

This study demonstrates that simplifying sample geometry by assuming that it matches the intended nominal shape when determining very soft tissue material properties through experiments and inverse analysis introduces a systematic bias, leading to appreciable underestimation of the material stress parameter. Models based on simplified geometries fail to accurately capture the kinematics of contact interactions between the sample's top surface and the loading platen of the testing machine, and they do not reproduce the complex, non-symmetric strain distributions that arise from such interactions and geometric imperfections in real samples. Consequently, although optimisation algorithms used in inverse analysis can compensate for these inaccuracies by adjusting stress parameters so that simplified-geometry models reproduce the experimental force–displacement curves, the resulting stress parameters differ substantially from those obtained using models based on the real (as determined from radiographic images) sample geometry. In this study, the largest discrepancy, nearly 50%, occurred for the shear modulus under compressive loading.

Therefore, our results suggest the importance of incorporating high-fidelity representations of tissue sample geometry into protocols for mechanical characterisation of the brain and other very soft tissues. An alternative or complementary strategy may involve developing excision or cutting techniques that constrain the differences between the actual and intended sample geometries.

**Acknowledgements** The authors acknowledge the facilities and scientific and technical assistance of the Western Australia National Imaging Facility (WA NIF) and thank Dr Tim Rosenow and Dr Sjoerd B. Vos of WA NIF and The University of Western Australia (UWA) for their support with image acquisition. The authors also thank Associate Professor Elena Juan Pardo, Head of the Translational 3D Printing Laboratory for Advanced Tissue Engineering (T3mPLATE) at the Harry Perkins Institute of Medical Research (Nedlands, WA, Australia), and the T3mPLATE laboratory personnel for providing access to the experimental equipment. The authors acknowledge Associate Professor Stuart Hodgetts (Spinal Cord Repair Laboratory, School of Human Sciences, The University of Western Australia) for his advice on sample dissection and storage, and Ms Ulani Hayter Otaola, a student at the School of Engineering, The University of Western Australia, for her assistance with conducting the experiments.

**Funding** This research was supported by the Australian Government through the Australian Research Council's Discovery Projects funding scheme (Project DP230100949). Author S. A. acknowledges support of the University of Western Australia Scholarship for International Research Fees and University of Western Australia (UWA) - Intelligent Systems for Medicine Laboratory (ISML) Higher Degree by Research Scholarship in Computational Biomechanics (partly funded from Australian Research Council's ARC Discovery Project DP230100949).